\documentclass[preprint,10pt,nofootinbib]{revtex4}
\usepackage[paper=letterpaper,margin=1.5in]{geometry}
\usepackage{epsfig,color}
\usepackage{graphicx}
\usepackage{amssymb,amsmath}
\newcommand{\be}{\begin{equation}}
\newcommand{\ee}{\end{equation}}

\begin{document}

\bibliographystyle{revtex}

\begin{flushright}
{\normalsize
SLAC-PUB-12370\\
DESY-07-023\\
February 2007}
\end{flushright}

\vspace{.4cm}

\title{Impedance Calculations of Non-Axisymmetric\\ Transitions
Using the Optical Approximation\footnote{Work supported by
Department of Energy contract DE--AC02--76SF00515 and and by the EU
contract 011935 EUROFEL.}}
\author{K.L.F.~Bane, G. Stupakov}
\affiliation{Stanford Linear Accelerator Center, Stanford
University, Stanford, CA 94309}
\author{I. Zagorodnov}
\affiliation{Deutsches Elektronen-Synchrotron, Notkestrasse 85,
22603 Hamburg, Germany\vspace{.4cm} }

\vspace{.4cm} \begin{abstract} In a companion report, we have
derived a method for finding the impedance at high frequencies of
vacuum chamber transitions that are short compared to the catch-up
distance, in a frequency regime that---in analogy to geometric
optics for light---we call the optical regime. In this report we
apply the method to various non-axisymmetric geometries such as
irises/short collimators in a beam pipe, step-in transitions,
step-out transitions, and more complicated transitions of practical
importance. Most of our results are analytical, with a few given in
terms of a simple one dimensional integral. Our results are compared
to wakefield simulations with the time-domain, finite-difference
program ECHO, and excellent agreement is found.  \vfill \centerline
{Submitted to Physical Review Special Topics--Accelerators and
Beams}
\end{abstract}

\maketitle


\section{Introduction}

In many current and future accelerator projects short, intense
bunches of charged particles are transported through vacuum chambers
that include objects such as transitions, irises, and collimators.
For example, in the beam delivery system of the International Linear
Collider (ILC) 3~nC, 300~$\mu$m-long bunches encounter many
collimators on their way to the interaction region~\cite{ILC}. Or,
in the undulator region of the Linac Coherent Light Source (LCLS) a
1~nC, 20~$\mu$m-long bunch passes by square-to-round transitions,
bpm's, and other changes in chamber geometry~\cite{LCLS}. Wakefields
generated by such changes in vacuum chamber geometry can negatively
affect the beam emittance and ultimately the performance of an
accelerator. Numerically obtaining the strength of the wakefields
or, equivalently, of the impedances for short bunches in such vacuum
chamber objects can be difficult, in particular when the object is
long and non-cylindrically symmetric.

In a recent paper~\cite{Stupakov07} we developed a method to solve
such seemingly difficult problems in a so-called \emph{optical
approximation}. This method is valid in the limit of high
frequencies and reduces the calculation of the impedance to the two
dimensional integration of potential functions. In this report we
make use of this method to work out solutions to a selection of 3D
(shorthand for non-cylindrically symmetric) geometries that can be
encountered in today's accelerators.

\begin{figure}[htbp]
\centering
\includegraphics*[height=50mm]{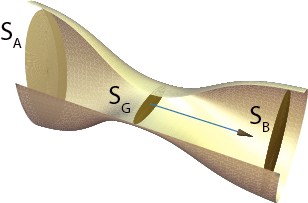}
\caption{Sketch of a generalized 3D transition, showing the regions
$A$, $G$, and $B$, with cross-sections $S_A$, $S_G$, and $S_B$,
respectively.}\label{transition_general_fi}
\end{figure}

The geometry of the problems to be considered is, in general, of the
type sketched in Fig.~\ref{transition_general_fi}. An in-going beam
pipe (region $A$ with cross-section profile $S_A$) is followed by a
short transition (the gap or aperture region $G$ with cross-section
$S_G$) and ends in an out-going pipe (region $B$ with cross-section
$S_B$).
We limit consideration to cases where $S_G$
is contained within the intersection of $S_A$ and $S_B$
[$S_G\subseteq (S_A\cap S_B)$]. The transition need not be smooth
like the one in the figure. A speed of light beam passes through the
three regions on a straight line that we call the {\it design
orbit}. At first there is no assumption on transverse symmetry in
the three region. However, we do assume that the axes of regions $A$
and $B$ are parallel to the design orbit. Note that if $S_{G}=S_B$
the structure is called a {\it step-in} transition (by which we mean
a short, inward transition), if $S_{G}=S_A$ it is a {\it step-out}
transition (a short, outward transition). If $S_A=S_B$, with the
aperture of $S_G$ smaller than the beam pipes, then the structure is
an iris (a metallic diaphragm with a hole or slot) or short
collimator in a beam pipe.

We are interested in finding the high frequency longitudinal or
transverse impedance of a general transition such as is sketched in
Fig.~\ref{transition_general_fi}. The impedance regime can be
described as the regime of {\it geometric optics}. This regime is
applicable provided that (1)~the frequency is high and (2)~the
transition is short. By condition~(1), we mean that the frequency
$\omega\gg c/g$, with $c$ the speed of light and $g$ the minimum
aperture of the structure. If the transition is tapered, we require
further that $\omega\gg c/(g\theta)$, with $\theta$ the taper angle.
By condition~(2), we mean that the length of the transition is short
compared to the catch-up distance, $\ell= g^2\omega/c$. Note that in
the optical regime the high frequency longitudinal impedance of a
structure is resistive, with the impedance $Z_\parallel$ real and
independent of frequency. The transverse impedance $Z_\perp$ is also
real and depends on frequency as $\omega^{-1}$.

The method of Ref.~\cite{Stupakov07} can be used to find the
impedance, for example, for an iris in a beam pipe, a step-in or
step-out transition, and a short collimator. Also, there are objects
that we call {\it long collimators}; {\it i.e.} collimators that
have a fixed (minimum) aperture over a length that is long compared
to the catch-up distance $\ell$, with transitions at the front and
back ends. For long collimators our methods also apply, provided
that the two transitions satisfy the above two conditions. For such
a collimator the impedance is the sum of the impedances of its two
transitions. Note, however, that for intermediate length
collimators, those of length comparable to the catch-up distance,
our methods do not apply.

In Ref.~\cite{Stupakov07} the impedance of a general transition in
the optical regime is developed following a systematic approach.
Earlier work on the subject was focused on specific geometries and
often was rather informal in nature. Balakin and Novokhatski were
among the first to address the question of impedance in the optical
regime \cite{Novokhatski}. Heifets and Kheifets studied the
longitudinal impedance of round step-in and step-out transitions (in
the optical regime) numerically by field matching~\cite{Heifets}.
For round, long collimators the dipole mode impedance was obtained
rigorously in Ref.~\cite{Palumbo}, and the higher azimuthal mode
impedances, more informally, in Ref.~\cite{Zimmermann}. The
impedance of short and long, round and 3D collimators was studied
numerically in Refs.~\cite{Zagorodnov06}, \cite{BaneZ06}; it was
found, for example, that the impedances of short and long
collimators, in general, differ significantly in amplitude (by a
factor $\sim2$ in the round, transverse case).

The impedances that we obtain in this report following our method
will also be compared to numerical results obtained by the
computer program ECHO~\cite{ECHO}. This program solves Maxwell's
equations to find the wakefield of an ultra-relativistic Gaussian
bunch within (perfectly conducting) metallic boundaries of fully
3D geometry. From the wakefields of a sufficiently short bunch the
impedance in the optical regime can be obtained directly. In the
present report discussions of ECHO calculations will be brief;
their main purpose is to confirm our results and to give us
confidence in our method.

This report is organized as follows: In Section~II the method of
calculation, derived in the companion paper, Ref.~\cite{Stupakov07},
is presented. At the end of this section some details of the ECHO
simulations will be given. The heart of the present report, however,
is the next four sections where our method is applied to 3D
transition examples. The problems that we solve are example irises
or short collimators within a beam pipe (Section~III), step-in
transitions (Section~IV), step-out transitions (Section~V), and more
complicated transitions (Section~VI). In these sections our results
will be compared briefly to ECHO simulation results. Section~VII
gives the conclusions. In the Appendix specific limits for the
impedance of an elliptical step-out transition are derived. Note
that although we follow the method of Ref.~\cite{Stupakov07} in this
report, the notation used here is not exactly the same. Note also
that Gaussian units will be used throughout; to convert impedances
to MKS units, one multiplies by the factor $Z_0c/4\pi$, with
$Z_0=377$~$\Omega$.

\section{Impedance Calculations}

\begin{figure}[htbp]
\centering
\includegraphics*[height=35mm]{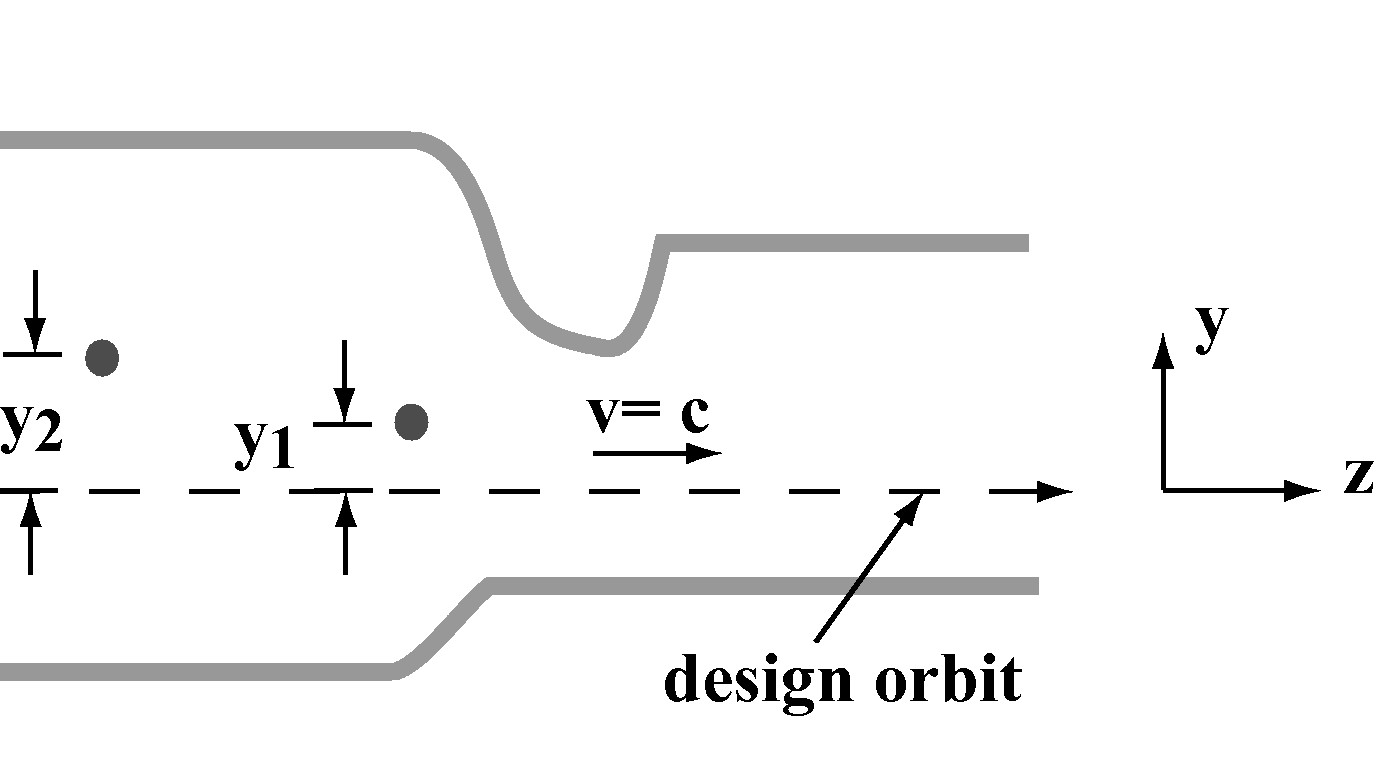}
\caption{Sketch of a transition showing the relationship of the
design trajectory and the vertical offsets of particle 1, $y_1$, and
particle 2, $y_2$ (the particles are indicated by red dots). For the
calculations the offsets $y_1$, $y_2$, are assumed to be
small.}\label{y1y2_sketch_fi}
\end{figure}

Consider a general (transversely non-symmetric) transition. Let the
design orbit follow the $z$-axis, with the particle motion in the
$+z$ direction. The transverse impedance of a general transition
consists of monopole, dipole, and quad\-ru\-pole components (with
respect to the design orbit) that involve tensors, and the method of
Ref.~\cite{Stupakov07} can, in principle, deal with such problems.
However, to simplify, in this report we will limit consideration to
geometries for which the design orbit lies on a vertical symmetry
plane of the boundaries, defined by (horizontal coordinate) $x=0$.
Then the transverse (vertical, $y$) impedance (with respect to the
reference trajectory) can be written as
\begin{equation}
Z_{\perp,tot}=Z_{\perp,m}+y_1 Z_{\perp,d}+ y_2 Z_{\perp,q}\
,\label{zperp_tot_eq}
\end{equation}
where the three terms are called the (transverse) monopole, dipole,
and quad\-ru\-pole contributions; where $y_1$ is a small offset of
particle~1 (the leading particle), and $y_2$ a small offset of
particle~2 (the trailing particle, see Fig.~\ref{y1y2_sketch_fi}).
Note that the horizontal impedance has an equation equivalent to
Eq.~(\ref{zperp_tot_eq}), with the quadrupole term equal to
$(-Z_{\perp,q})$.

For most of the examples of this paper, the system has also a
horizontal symmetry plane and the design orbit lies in this plane.
In such a case $Z_{\perp,m}=0$; if, in addition, $y_1=y_2=y_0$ we
can define a normalized total impedance
\begin{equation}
Z_{\perp}=\frac{Z_{\perp,tot}}{y_0}=Z_{\perp,d}+Z_{\perp,q}\ .
\end{equation}
By convention, this is the normal definition of total transverse
impedance for bi-symmetric problems. (In this paper, however, we
will calculate the individual terms $Z_{\perp,d}$ and
$Z_{\perp,q}$ as well as $Z_{\perp}$ for bi-symmetric problems.)
The transverse wake at position $s$ within a bunch ($s<0$ is
toward the head), normalized to the offset $y_0$, in the optical
regime is given by
\begin{equation}
W_\perp(s)=\frac{i}{2\pi}\int_{-\infty}^{\infty}Z_\perp(\omega){\tilde
\lambda_z}(\omega)e^{-i\omega s/c}\,d\omega=(\omega
Z_\perp)\int_{-\infty}^s\lambda_z(s')\,ds'\ ,
\end{equation}
with ${\tilde \lambda_z}(\omega)$ the Fourier transform of the
line charge density $\lambda_z(s)$. To arrive at the last
expression in this equation we have used the fact that, in the
optical regime, $Z_\perp\propto\omega^{-1}$. Thus the quantity
$(\omega Z_\perp)$ is independent of frequency. The result is
proportional to the integral of the line charge density (and in
the longitudinal case, it is proportional to the line charge
density itself). The kick factor $\kappa_\perp$, the average kick
experienced by the beam per unit charge per unit offset, is the
integral of $W_\perp(s)$ when weighted by the longitudinal charge
density. In the optical regime the kick factor is simply given by
the constant
\begin{equation}
\kappa_\perp=\frac{(\omega Z_\perp)}{2}\ .\label{kappa_eq}
\end{equation}

In Ref.~\cite{Stupakov07} using the Panofsky-Wenzel
theorem~\cite{Panofsky56}, the transverse monopole, dipole, and
quadrupole terms are related to longitudinal monopole, dipole, and
quadrupole impedances:
\begin{equation}
Z_{\perp,m}=\frac{c}{\omega} Z_{\parallel,m}\ ,\quad\quad
Z_{\perp,d}=\frac{c}{\omega }Z_{\parallel,d}\ ,\quad\quad
Z_{\perp,q}=\frac{2c}{\omega }Z_{\parallel,q}\
.\label{zperp_zparallel_eq}
\end{equation}
The longitudinal impedances can, in turn, be obtained from
integrals involving the Green functions to Poisson's equation (the
potentials) in regions $A$ and $B$ of the transition of interest.
In the case of the monopole part
\begin{equation}
Z_{\parallel,m} =\frac{1}{2\pi
c}\left[\int_{S_B}\nabla\varphi_{m,B}\cdot\nabla\varphi_{d,B}\,dS
-\int_{S_{G}}\nabla\varphi_{m,A}\cdot\nabla\varphi_{d,B} \,dS
\right]\ ,\label{Rm_eq}
\end{equation}
with the integrals taken over the cross-section areas $S_B$ and
$S_G$, respectively. Here the monopole and quad potentials in region
$A$ are given by the solutions to
\begin{equation}
\nabla^2\varphi_{m,A}= -4\pi \delta(y)\delta(x)\ ,\quad\quad
\nabla^2\varphi_{d,A}= 4\pi \delta'(y)\delta(x)\
,\label{laplacemd_eq}
\end{equation}
with boundary conditions $\varphi_{m,A}=0$, $\varphi_{d,A}=0$, on
metallic boundary $C_A$ that encloses $S_A$. Similar equations hold
for $\varphi_{m,B}$ and $\varphi_{d,B}$ of region $B$.

In the case of the dipole part of the impedance
\begin{equation}
Z_{\parallel,d} =\frac{1}{2\pi
c}\left[\int_{S_B}(\nabla\varphi_{d,B})^2\,dS
-\int_{S_{G}}\nabla\varphi_{d,A}\cdot\nabla\varphi_{d,B}\,dS
\right]\ .\label{Rd_eq}
\end{equation}
For the quad part of impedance
\begin{equation}
Z_{\parallel,q} =\frac{1}{2\pi
c}\left[\int_{S_B}\nabla\varphi_{m,B}\cdot\nabla\varphi_{q,B}\,dS
-\int_{S_{G}}\nabla\varphi_{m,A}\cdot\nabla\varphi_{q,B} \,dS
\right]\ .\label{Rq_eq}
\end{equation}
The quadrupole potential in region~$B$ is given by the solution to
\begin{equation}
\nabla^2\varphi_{q,B}= -2\pi \delta''(y)\delta(x)\
,\label{laplaceq_eq}
\end{equation}
with $\varphi_{q,B}=0$ on boundary $C_B$. Finally, the
longitudinal impedance on the reference trajectory is given by:
\begin{equation}
Z_{\parallel,long} =\frac{1}{2\pi
c}\left[\int_{S_B}(\nabla\varphi_{m,B})^2\,dS
-\int_{S_{G}}\nabla\varphi_{m,A}\cdot\nabla\varphi_{m,B}\,dS
\right]\ .\label{Rlong_eq}
\end{equation}

To obtain the needed potentials $\varphi_m$, $\varphi_d$,
$\varphi_q$, we begin with the Green function solution to
\begin{equation}
\nabla^2G(x,y,y_0)=-4\pi\delta(x)\delta(y-y_0)\ ,
\end{equation}
where $G=0$ on the boundary $C$ of the surface of interest $S$. Here
the source particle is located at $x=0$, $y=y_0$. For region~$S$,
$\varphi_{m}(x,y)=G(x,y,0)$,
\begin{equation}
\varphi_{d}(x,y)=\left[\frac{\partial}{\partial
y_0}G(x,y,y_0)\right]_{y_0=0}\ ,\quad\
\varphi_{q}(x,y)=\frac{1}{2}\left[\frac{\partial^2}{\partial
y_{0}^2}G(x,y,y_0)\right]_{y_0=0}\ .\label{varphi_eq}
\end{equation}

As was shown in Ref.~\cite{Stupakov07}, with the help of Green's
first identity~\cite{Green},
\begin{equation}
\int_S(\phi\nabla^2\psi+\nabla\phi\cdot\nabla\psi)\,dS=
\int_C\phi{\bf n}\cdot{\mathbf\nabla}\psi\,dl\ ,
\end{equation}
the surface integrals involved in the impedance equations can be
converted to line integrals. In this formula $\phi$ and $\psi$ are
functions that can be differentiated twice; $S$ is a surface over
which a surface integration is performed; $C$ is the contour that
encloses $S$ and over which a line integral is performed; and ${\bf
n}$ is a unit normal vector pointing outward from the contour. This
device is used throughout the examples of this report.

\subsection*{Numerical Comparisons}

A great number of ECHO simulations were performed to test and
verify our results. ECHO is a 3D, time-domain finite difference
program that calculates wakefields generated by an
ultra-relativistic bunch passing through the structure. ECHO has
two features that make these 3D, optical regime simulations
tractable. (1)~A method to reduce the so-called ``mesh
dispersion''---errors generated in time-domain mesh programs that
are especially difficult to deal with for the combination of short
bunches and long structures. (2)~An indirect method of calculating
wakes in 3D structures that eliminates long downstream beam pipes,
which would be needed in the case of direct wake calculation with
very short bunches. Since the ECHO simulations play a supporting
role in the present report, other than comparing results, we will
only discuss them briefly. For a more detailed report on ECHO, see
{\it e.g.} Ref.~\cite{ECHO}.

A few comments on the parameters used in the ECHO simulations: The
bunches in the simulations are Gaussian. Typically we choose
$\sigma_z=g/10$, with $g$ the minimum vertical half aperture of the
iris or transition. For a flat iris or beam pipe we take the
structure half-width $w=10g$. (In this report we use the word {\it
flat} to mean ``having a rectangular cross-section with a small
height to width ratio.'') For a small iris in a beam pipe or a
transition opening into a large beam pipe, the beam pipe is square
with half-height $10g$. We take mesh sizes to be
$\frac{1}{5}\sigma_z$, $\frac{5}{8}\sigma_z$, $\frac{5}{8}\sigma_z$,
in the longitudinal, horizontal, vertical directions, respectively.
Irises have the thickness of one longitudinal mesh size. For
bi-symmetric structures ECHO calculates the dipole and quadrupole
components of the wakes independently.

\begin{figure}[htbp]
\centering
\includegraphics*[height=55mm]{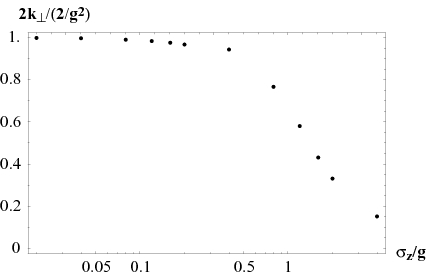}
\caption{Twice the kick factor $k_\perp$ of a thin round iris in a
beam pipe {\it vs.} rms bunch length $\sigma_z$, as obtained by
ECHO. The iris radius is $g$.}\label{igors_iris_wakes_summary_fi}
\end{figure}

\begin{figure}[htbp]
\centering
\includegraphics*[height=50mm]{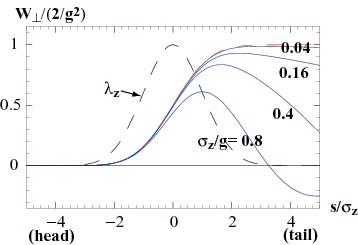}
\caption{Transverse wake of a thin round iris in a beam pipe as
obtained by ECHO. Results for bunch lengths over iris radius
$\sigma_z/g= 0.04$, 0.16, 0.4, and 0.8 are given. The (Gaussian)
bunch shape, with the head to the left, is indicated by the black
dashes. The analytical wake of a Gaussian bunch in a round iris in
the optical regime, $W_\perp=[1+{\rm erf}(s/\sqrt{2}\sigma_z)]/g^2$,
with ${\rm erf}(x)$ the error function, is also given (the red
dashes).}\label{igors_iris_wakes_fi}
\end{figure}

To demonstrate the validity of the optical approximation, and how
one moves out of the optical regime as the bunch length increases,
we consider the problem of a thin round iris of radius $g$ in a
round beam pipe of radius $b$. We choose $b/g=4$. We perform ECHO
calculations for bunch lengths in the range $\sigma_z/g=0.02$ to 4.
In Fig.~\ref{igors_iris_wakes_summary_fi} we plot twice the kick
factor of our results as function of bunch length. According to
Eq.~\ref{kappa_eq}, in the optical regime $2k_\perp=\omega Z_\perp$,
which is a number independent of bunch length or frequency. We see
that the result of our numerical calculation is fairly constant up
until $\sigma_z/g\sim0.2$, after which the validity of the optical
approximation starts to break down. In more detail, we give in
Fig.~\ref{igors_iris_wakes_fi} the actual wake functions for four
example bunch lengths as obtained by ECHO. The (Gaussian) bunch
shape, with the head to the left, is indicated by the black dashes.
The analytical wake of a Gaussian bunch in a round iris in the
optical regime, $W_\perp=[1+{\rm erf}(s/\sqrt{2}\sigma_z)]/g^2$,
with ${\rm erf}(x)$ the error function, is also given (the red
dashes). We see that, although the weighted average of the wake
agrees well for $\sigma_z/g\sim0.2$, good agreement between the
wakes over $\pm4\sigma_z$ is not obtained until $\sigma_z/g\sim
0.04$.

\section{Iris/Short Collimator in Beam Pipe}\label{sec:Short_Collimator}

\begin{figure}[htbp]
\centering
\includegraphics*[height=65mm]{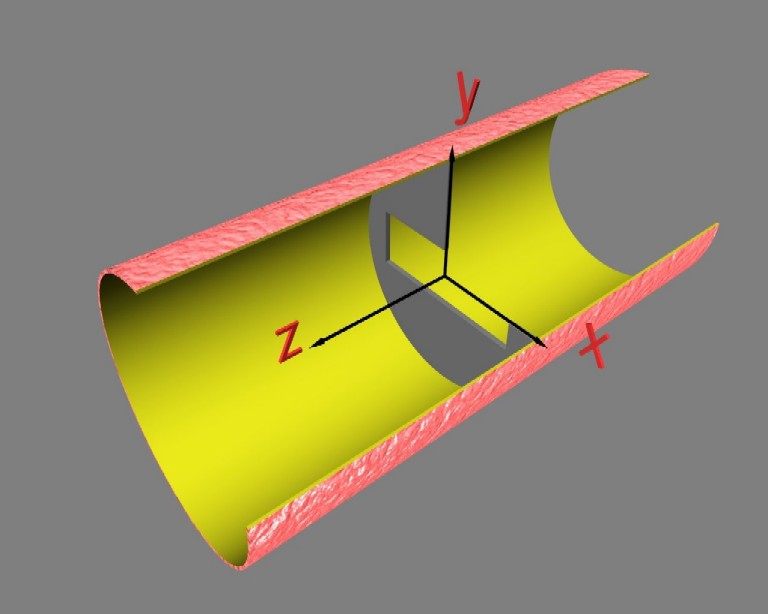}
\caption{An iris in a beam pipe.}\label{iris_sketch_fi}
\end{figure}

We begin by performing calculations of examples with thin irises or
short collimators in a beam pipe (for an example, see
Fig.~\ref{iris_sketch_fi}). In these cases the solution is given by
a particularly simple ``clipping'' type calculation of the energy
impinging on the iris wall.

For these examples the beam pipes in Region~A and B are identical,
{\it i.e.} $S_B=S_A$. Thus Eqs.~(\ref{zperp_zparallel_eq}),
(\ref{Rd_eq}), (\ref{Rq_eq}), give
\begin{equation}
Z_{\perp,d} =\frac{1}{2\pi \omega
}\int_{S_B-S_G}(\nabla\varphi_{d,B})^2\,dS
 \
,\label{Rd_thin_eq}
\end{equation}
\begin{equation}
Z_{\perp,q} =\frac{1}{\pi \omega
}\int_{S_B-S_G}\nabla\varphi_{m,B}\cdot\nabla\varphi_{q,B}\,dS
 \
.\label{Rq_thin_eq}
\end{equation}
Both equations involve integrals over the metallic surface of the
iris. Using Green's first identity and considering the fact that
$\nabla\varphi^2=0$ in any region that precludes the axis, we can
write Eqs.~(\ref{Rd_thin_eq}), (\ref{Rq_thin_eq}), as one
dimensional integrals. For example, Eq.~(\ref{Rd_thin_eq}) becomes
\begin{equation}
Z_{\perp,d} =\frac{1}{2\pi \omega }\int_{C_G}\varphi_{d,B}{\bf
n}\cdot\nabla\varphi_{d,B}\,dl\ ,\label{Rd_thinb_eq}
\end{equation}
where the integral on the right is a line integral over the curve
that encloses the area of $S_G$, which we denote by $C_G$, and ${\bf
n}$ is a unit vector normal to this curve in the direction of the
axis (an identical integral over $C_B$ was dropped, since
$\varphi_{d,B}=0$ on $C_B$). Similarly,
\begin{equation}
Z_{\perp,q} =\frac{1}{\pi \omega }\int_{C_G}\varphi_{m,B}{\bf
n}\cdot\nabla\varphi_{q,B}\,dl\ .\label{Rq_thinb_eq}
\end{equation}

\begin{figure}[htbp]
\centering
\includegraphics*[width=95mm]{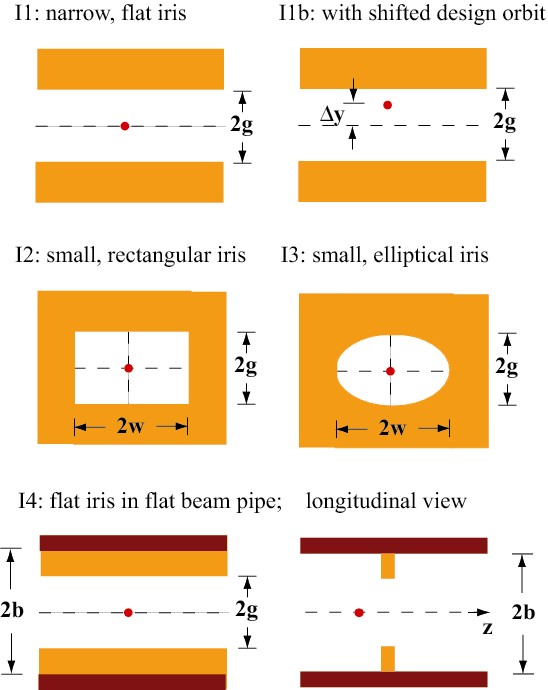}
\caption{Cross-section views of the iris/short collimator geometries
considered in this section (for case I4 a longitudinal view is also
given). Dimension labels are given, and the design orbit location is
indicated by the red dot. In cases I1-I3 the beam pipe aperture is
specified as large compared to the iris aperture; in case I4 the
beam pipe is flat, with vertical aperture $2b$. For case I4 two
colors are used for the boundaries as an aid in
visualization.}\label{iris_geometries_fi}
\end{figure}

The examples dealt with in this section are: (I1)~a small, flat
iris (or horizontal slot) in a beam pipe, (I1b)~the same but with
the design orbit shifted vertically, (I2)~a small rectangular
iris, (I3)~a small elliptical iris, (I4)~a flat iris (not
necessarily small) in a flat beam pipe. In all cases except I1b,
the design orbit is on a horizontal symmetry plane. A
cross-section view of the geometries is sketched in
Fig.~\ref{iris_geometries_fi}. Dimension labels and the design
orbit location are also shown.

In the following subsections (except for subsection I4) we will
assume that the pipe radius is very large compared to the iris
aperture. We can use the free-space Green function for a particle
vertically offset by $y_0$:
\begin{equation}
G(x,y,y_0)= -\ln[x^2+(y-y_0)^2]\ .\label{green_free_eq}
\end{equation}
From this Green function we obtain the potentials:
\begin{equation}
\varphi_m=-\ln(x^2+y^2)\ ,\quad\varphi_d=\frac{2y}{x^2+y^2}\
,\quad\varphi_q=\frac{y^2-x^2}{(x^2+y^2)^2}\ .
\end{equation}
Then using these potentials, we perform the calculations of
Eqs.~(\ref{Rd_thinb_eq}), (\ref{Rq_thinb_eq}).

The error in transverse impedance introduced by the approximation of
the free-space Green function is relatively small. For example, in
the case of a round ({\it i.e.} cylindrically symmetric) iris of
radius $g$ within a round beam pipe of radius $b$ it is known that
the correction to the first order term is $(g/b)^4$
\cite{Zagorodnov06}.

\subsection*{I1. Flat Iris or Horizontal Slot in Large Pipe}

Consider a large beam pipe containing an iris with a horizontal slot
of vertical size $2g$, with $g$ very small compared to the beam pipe
size. We take the reference trajectory to be the symmetry line, so
the monopole term $Z_{\perp,m}=0$. The dipole term is given by
\begin{equation}\label{Z_perpend}
Z_{\perp,d} =-\frac{2}{\pi \omega
}\int_0^\infty\varphi_{d,B}(x,g)\frac{\partial\varphi_{d,B}}{\partial
y}(x,g)\,dx
 \ ,
\end{equation}
and similarly for the quadrupole term. With the free-space
potentials the integrals can be performed analytically, and after
some algebra we obtain
\begin{equation}
Z_{\perp,d}= Z_{\perp,q}= \frac{1}{\omega g^2}\ ,\quad Z_{\perp}=
\frac{2}{\omega g^2}\ .
\end{equation}
We see that the dipole and quad terms are equal, and that the
total result is the same as the leading order impedance for a {\it
round}, thin iris of radius $g$~\cite{Zagorodnov06}. Note that for
a flat iris that is oriented vertically, we have $Z_{\perp,d}=
-Z_{\perp,q}= 1/(\omega g^2)$ and $Z_{\perp}=0$.

\subsubsection*{I1b. Case of Shifted Design Orbit}
We can also find the impedance for the case the design orbit is
shifted vertically by $\Delta y$. In this case the (transverse)
monopole term $Z_{\perp,m}$ dominates the transverse impedance, and
we can neglect the effect of $Z_{\perp,d}$ and $Z_{\perp,q}$ [see
Eq.~(\ref{zperp_tot_eq})]. The transverse monopole term is obtained
using Eqs.~(\ref{zperp_zparallel_eq}), (\ref{Rm_eq}). Then the
calculation proceeds in the same manner as before:
\begin{eqnarray}
Z_{\perp,m}&=&-\frac{1}{\pi\omega}\int_0^\infty\left[\varphi_{d,B}(x,g-\Delta
y)\frac{\partial\varphi_{m,B}}{\partial y}(x,g-\Delta y)\right.\\
\nonumber & &\quad\quad\quad+\left. \varphi_{d,B}(x,-g-\Delta
y)\frac{\partial\varphi_{m,B}}{\partial y}(x,-g-\Delta
y)\right]\,dx\ .\label{zd_iris_eq}
\end{eqnarray}
We find that the transverse impedance is given by
\begin{equation}
Z_{\perp,m}= \frac{1}{\omega}\left[\frac{1}{g-\Delta
y}-\frac{1}{g+\Delta y}\right]\ .\label{zm_iris_eq}
\end{equation}
For small $\Delta y$, $Z_{\perp,m}=2\Delta y/(\omega g^2)$, which is
consistent with our earlier results.

\subsection*{I2. Rectangular Iris}

For the case of a rectangular iris with a small aperture of $2w$ by
$2g$ (horizontal by vertical) the calculation follows the same
procedure as before; however, the integration now is a line integral
along the rectangular aperture. We perform integrals like
\begin{equation}
Z_{\perp,d} =-\frac{2}{\pi \omega
}\left[\int_0^w\varphi_{d,B}(x,g)\frac{\partial\varphi_{d,B}}{\partial
y}(x,g)\,dx
+\int_0^g\varphi_{d,B}(w,y)\frac{\partial\varphi_{d,B}}{\partial
x}(w,y)\,dy\right]
 \ .
\end{equation}
The final solution is
\begin{eqnarray}
Z_{\perp,d}&=&\frac{2}{\pi\omega g^2}\frac{\alpha+{\rm
arccot}(\alpha)+\alpha^2\arctan(\alpha)}{\alpha^2}\nonumber\\
Z_{\perp,q}&=&\frac{2}{\pi\omega
g^2}\frac{\alpha(-1+\alpha^2)+(1+\alpha^2)[{-\rm
arccot}(\alpha)+\alpha^2\arctan(\alpha)]}{\alpha^2(1+\alpha^2)}\nonumber\\
Z_\perp&=&\frac{4}{\pi\omega
g^2}\frac{\alpha+(1+\alpha^2)\arctan(\alpha)}{1+\alpha^2}\label{rect_iris_eq}
\end{eqnarray}
where $\alpha=w/g$.
\begin{figure}[htbp]
\centering
\includegraphics*[height=50mm]{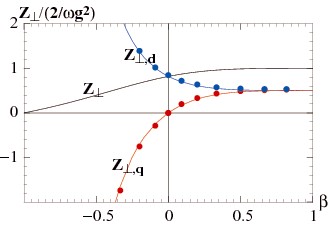}
\caption{For a rectangular iris in a beam pipe, the transverse
impedances $Z_{\perp}$, $Z_{\perp,d}$, $Z_{\perp,q}$, normalized to
$2/(\omega g^2)$, as functions of $\beta=(w-g)/(w+g)$. Plotting
symbols give ECHO numerical results.}\label{rect_iris_fi}
\end{figure}

The results, when normalized to $2/\omega g^2$, are plotted as
functions of $\beta=(w-g)/(w+g)$ in Fig.~\ref{rect_iris_fi}. Note
that when $\beta= 1$ the results agree with the results for the iris
with the infinitely wide horizontal slot, given above. For
$\beta=-1$ (an infinitely high vertical slot)
$Z_{\perp,q}=-Z_{\perp,d}=-1/(\omega w^2)$ and $Z_{\perp}=0$. For
the special case of a square aperture
$Z_{\perp}=Z_{\perp,d}=2(\frac{1}{\pi}+\frac{1}{2})/g^2$ and
$Z_{\perp,q}=0$. Note that the {\it horizontal} impedance is
obtained from Eqs.~(\ref{rect_iris_eq}) by exchanging $w$ and $g$.

In Fig.~\ref{rect_iris_fi} ECHO numerical results for $Z_{\perp,d}$
and $Z_{\perp,q}$ are shown by the plotting symbols. We see good
agreement with the results of our method.

\subsection*{I3. Elliptical Iris}

\begin{figure}[htbp]
\centering
\includegraphics*[height=50mm]{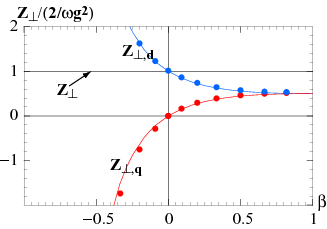}
\caption{For an elliptical iris in a beam pipe, the transverse
impedances $Z_{\perp}$, $Z_{\perp,d}$, $Z_{\perp,q}$, normalized to
$2/(\omega g^2)$, as functions of $\beta=(w-g)/(w+g)$. Plotting
symbols give ECHO numerical results.}\label{ellipse_iris_fi}
\end{figure}

The impedance calculation for an elliptical iris with a small
aperture, with axes $w$ by $g$ (horizontal by vertical), follows in
a similar manner to the rectangular iris case. The equations that
need to be solved are Eqs.~(\ref{Rd_thinb_eq}), (\ref{Rq_thinb_eq}).
We see that the elliptical case is more complicated in that it
requires the formation of ${\bf n}\cdot\nabla\varphi_{q,B}$ and of
the length metric on the ellipse of the iris, though the final
solution is quite simple. We find that
\begin{equation}
Z_{\perp,d}=\frac{1}{\omega g^2}\left(1+\frac{g^2}{w^2}\right)\
,\quad Z_{\perp,q}=\frac{1}{\omega
g^2}\left(1-\frac{g^2}{w^2}\right)\ ,\quad Z_{\perp}=\frac{2}{\omega
g^2}\ .\label{ellipse_iris_eq}
\end{equation}
We find that for an elliptical iris $Z_\perp$ is independent of the
size of the horizontal axis $w$! As was true for the rectangular
iris, the horizontal impedance of the elliptical iris can be
obtained by exchanging $w$ and $g$ in the solution equations. In
Fig.~\ref{ellipse_iris_fi} we plot the analytical results and
compare with ECHO numerical results. We see good agreement.

\subsection*{I4. Flat Iris in Flat Beam Pipe}\label{sec:I4}

Consider a flat iris of height $2g$ centered within a flat beam pipe
of height $2b$. We begin this problem with the Green function
between two parallel plates of aperture $2b$~\cite{MorseFeshbach}
\begin{equation}
G(x,y,y_0)=\ln\left[\frac{\cosh\frac{\pi
x}{2b}+\cos\frac{\pi(y+y_0)}{2b} } {\cosh\frac{\pi x}
{2b}-\cos\frac{\pi(y-y_0)}{2b}}\right]\ .\label{green_parallel_eq}
\end{equation}
The calculation procedure is the same as for the
free-space-to-flat-iris example, Example~I1. The potentials are

\begin{eqnarray}
\varphi_{m,B}(x,y)&=&\ln\left[\frac{\cosh\frac{\pi
x}{2b}+\cos\frac{\pi y}{2b} } {\cosh\frac{\pi x} {2b}-\cos\frac{\pi
y}{2b}}\right]\nonumber\\
 \varphi_{d,B}(x,y)&=&\frac{\pi}{2b}\,\frac{\sin\frac{\pi
y}{b}} {\cosh^2\frac{\pi x}{2b}-\cos^2\frac{\pi y}{2b}}\nonumber\\
\varphi_{q,B }(x,y)&=&\frac{\pi^2}{8b^2}\,\frac{(-2+\cosh\frac{\pi
x}{b}+\cos\frac{\pi y}{b})\cosh\frac{\pi x}{2b}\cos\frac{\pi y}{2b}}
{\left(\cosh^2\frac{\pi x}{2b}-\cos^2\frac{\pi y}{2b}\right)^2}\ .
\label{fi_flat_to_flat_eq}
\end{eqnarray}

\begin{figure}[htbp]
\centering
\includegraphics*[height=50mm]{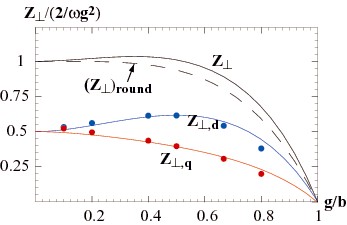}
\caption{For a flat iris with aperture $2g$ in a flat beam pipe of
aperture $2b$, the transverse impedances $Z_{\perp}$, $Z_{\perp,d}$,
$Z_{\perp,q}$ as functions of $\alpha=g/b$. Plotting symbols give
ECHO numerical results. The round case,
$(Z_\perp)_{round}=2(1/g^{2}-g^2/b^{4})/\omega$, is also shown
(dashes).}\label{flat_to_flat_thin_fi}
\end{figure}

The final solution is
\begin{eqnarray}
Z_{\perp,d}&=&\frac{\pi \alpha^2}{2\omega
g^2}\csc^2(\pi\alpha)[2\pi(1-\alpha)+\sin(2\pi\alpha)]\nonumber\\
Z_{\perp,q}&=&\frac{\pi \alpha^2}{\omega
g^2}\csc(\pi\alpha)[1+\pi(1-\alpha)\cot(\pi\alpha)]\nonumber\\
Z_\perp&=&\frac{\pi \alpha^2}{2\omega
g^2}\csc^2(\pi\alpha/2)[\pi(1-\alpha)+\sin(\pi\alpha)]
\end{eqnarray}
where $\alpha=g/b$. These curves are plotted in
Fig.~\ref{flat_to_flat_thin_fi}. The round case, with $g$ and $b$,
representing, respectively, the radii of the iris and of the beam
pipe,
$(Z_\perp)_{round}=2(1/g^{2}-g^2/b^{4})/\omega$~\cite{Zagorodnov06},
is also shown (the dashes). We note that $Z_\perp$ is always close
to and larger for the flat than for the round case.

In Fig.~\ref{flat_to_flat_thin_fi} ECHO numerical results for
$Z_{\perp,d}$ and $Z_{\perp,q}$ are again shown by plotting
symbols. We see basically good agreement with the results of our
method. We should point out one subtlety however. For $g/b= 0.8$
the numerically obtained wake begins to take on some inductive
character and the result for impedance begins to deviate from our
analytical solution (particularly $Z_{\perp,d}$). We attribute
this to the fact that, for the parameters used in the simulations,
$(b-g)/b=\sigma_z$; as $(b-g)/b\lesssim\sigma_z$ we are beginning
to leave the optical regime.

\subsubsection*{Longitudinal Impedance}

\begin{figure}[htbp]
\centering
\includegraphics*[height=50mm]{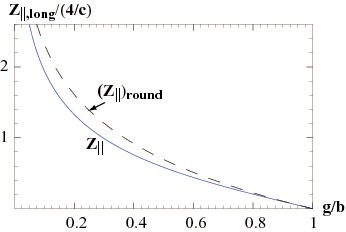}
\caption{For a thin flat iris with aperture $2g$ in a flat beam pipe
of aperture $2b$, the longitudinal impedance $Z_{\parallel,long}$ as
function of $\alpha=g/b$. The round case,
$(Z_{\parallel})_{round}=4\ln(b/g)/c$, is also shown
(dashes).}\label{flat_to_flat_thin_long_fi}
\end{figure}

If we know the geometry of both the beam pipe and the iris we can
also obtain the longitudinal impedance, $Z_{\parallel,long}$. For
the longitudinal impedance we begin with Eq.~(\ref{Rlong_eq}), which
we convert to a line integral, and obtain
\begin{eqnarray}
Z_{\parallel,long} &=&-\frac{2}{\pi
c}\int_0^\infty\varphi_{m,B}\frac{\partial\varphi_{m,B}}{\partial
y}(x,g)\,dx\nonumber\\ &=&\frac{8}{
c}\int_0^\infty\frac{\sin\pi\alpha\cosh\pi
x}{\cos2\pi\alpha-\cosh2\pi x}\ln\left(\frac{\cosh\pi
x-\cos\pi\alpha}{\cosh\pi x+\cos\pi\alpha}\right)\,dx\ ,
\end{eqnarray}
with $\alpha=g/b$. The last integral we evaluate numerically. The
result, when normalized to $4/c$, is shown in shown in
Fig.~\ref{flat_to_flat_thin_long_fi}. For comparison, the round case
impedance (with $g$ and $b$ representing, respectively, the iris and
beam pipe radius), is also shown. We see that the longitudinal
impedance in the flat case is always less than in the round case.

\section{Step-In Transition}

For the dipole component of impedance of a step-in transition we
replace, in Eq.~(\ref{Rd_eq}), $S_{G}$ by $S_B$, and use Green's
first identity to obtain
\begin{eqnarray}
Z_{\parallel,d} &=&\frac{1}{2\pi c}\left[\int_{C_B}\varphi_{d,B}{\bf
n}\cdot{\bf\nabla}\varphi_{d,B}\,dl
-\int_{S_{B}}\varphi_{d,B}\nabla^2\varphi_{d,B}\,dS\right.\nonumber\\
& &\quad\quad\left.-\int_{C_B}\varphi_{d,B}{\bf
n}\cdot{\bf\nabla}\varphi_{d,A}\,dl
+\int_{S_{B}}\varphi_{d,B}\nabla^2\varphi_{d,A}\,dS \right]=0
\end{eqnarray}
The first and third integrals are zero because $\varphi_{d,B}=0$ on
boundary $C_B$; the second and fourth integrals cancel because the
Laplacian is the same independent of region.

The same kind of analysis shows that $Z_{\parallel,q}=0$ in a
step-in transition, and that $Z_{\parallel,m}=0$ also in
non-symmetric step-in transitions. Since specifics of the geometry
were not used in our derivation, we conclude that the total
transverse impedance is zero for a (short) step-in transition of
{\it any} geometry. Similarly the {\it longitudinal} impedance of
any short  step-in transition is also zero in the optical regime
(something that has been known to be true for the special case of a
round step-in transition~\cite{Novokhatski}, \cite{Heifets}). We
conclude that the longitudinal and transverse impedance of {\it any}
step-in transition, in the optical regime, is zero!

The impedance of a long collimator is just the sum of the impedances
of a step-in and a step-out transition. Since the impedance of a
step-in transition is zero, the impedance of a long collimator is
the same as that of a step-out transition alone, the impedance of
which we will study in the following section.

\section{Step-Out Transition}

\begin{figure}[htbp]
\centering
\includegraphics*[height=65mm]{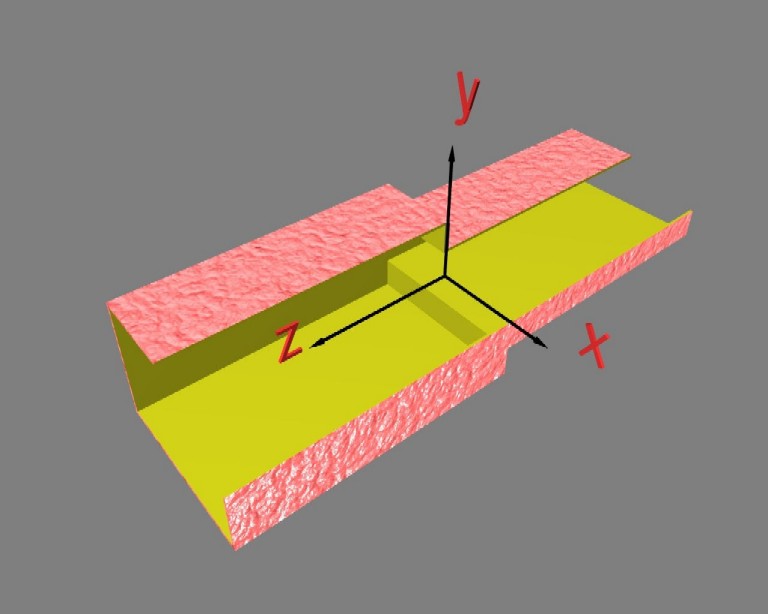}
\caption{A step-out transition. The particles move in the $+z$
direction.}\label{transition_sketch_fi}
\end{figure}

\begin{figure}[htbp]
\centering
\includegraphics*[width=90mm]{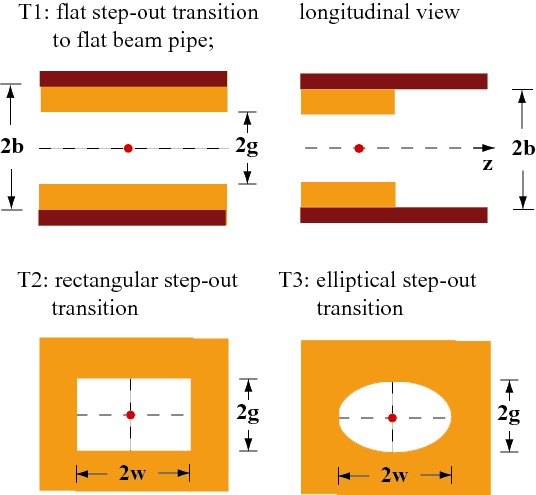}
\caption{Cross-section views of the step-out transition geometries
considered in this section (for case T1 a longitudinal view is also
given). Dimension labels are given, and the design orbit location is
indicated by the red dot. In cases T2-T3 the outgoing beam pipe
aperture is specified as large compared to the incoming beam pipe
aperture; in case T1 the outgoing beam pipe is flat, with vertical
aperture $2b$. For case T1 two colors are used for the boundaries as
an aid in visualization.}\label{trans_geometries_fi}
\end{figure}

An example step-out transition is sketched in
Fig.~\ref{transition_sketch_fi} (the particles move in the $+z$
direction). The examples dealt with in this section are: (T1)~a
flat beam pipe that transitions to larger flat beam pipe, (T2)~a
rectangular pipe that transitions to a large pipe, and (T3)~an
elliptical pipe that transitions to a large pipe. Cross-section
views of the geometries (and a longitudinal view in case (T1) are
sketched in Fig.~\ref{trans_geometries_fi}. Dimension labels and
the design orbit location are also shown.

For the impedance of a step-out transition we replace, in
Eq.~(\ref{Rd_eq}), $S_{G}$ by $S_A$. The impedance function (in the
dipole case) $Z_{\parallel,d}$ of a step-out transition can be
written, beginning with Eq.~(\ref{Rd_eq}), as
\begin{align}
Z_{\parallel,d} =\frac{1}{2\pi
c}&\left[\int_{S_B}(\nabla\varphi_{d,B})^2\,dS
-\int_{S_A}\nabla\varphi_{d,A}\cdot\nabla\varphi_{d,B}\,dS\right]\nonumber\\
=\frac{1}{2\pi
c}&\left[\int_{S_B}(\nabla\varphi_{d,B})^2\,dS-\int_{S_A}(\nabla\varphi_{d,A})^2\,dS\right.\nonumber\\
&\left.-\int_{S_{A}}\nabla\varphi_{d,A}\cdot\nabla(\varphi_{d,B}-\varphi_{d,A})\,dS
\right]\ .
\end{align}
By using Green's identity one can easily see that the last integral
is zero. Thus
\begin{equation}
Z_{\parallel,d} =\frac{1}{2\pi
c}\left[\int_{S_B}(\nabla\varphi_{d,B})^2\,dS-\int_{S_A}(\nabla\varphi_{d,A})^2\,dS
\right]\ .\label{Z_trans_edif_eq}
\end{equation}
This result can be interpreted as $2/c$ times the static field
energy of a dipole in region~$B$ minus that of a dipole in
region~$A$. This principle---that the longitudinal impedance (here
longitudinal dipole impedance) of a step-out transition in the
optical regime is given by twice the difference in the static field
energy in the two regions---was first elaborated for the round
step-out transition (longitudinal case) by Heifets and
Kheifets~\cite{Heifets}. The principle was then used for obtaining
higher order wakes (azimuthal mode number $m>0$) of (long) round
collimators~\cite{Zimmermann}, and also for specific 3D
structures~\cite{Zagorodnov06}.

We see that a main difference in the calculation of the impedance of
a step-out transition and of an iris is that, in the former case the
surface integrals are performed over a region that includes the
source charges of the potentials, and in the latter case the region
of the charges is excluded. In the step-out transition case,
however, it turns out that even though the potentials diverge at the
source charges, for the impedance two terms with the same divergence
are subtracted, and the impedance is finite.

By regrouping integrals and using Green's identity we can write the
impedance of a step-out transition in terms of a simple line
integral. The impedance function (for example, in the dipole case)
can be written, beginning with Eq.~(\ref{Rd_eq}), as
\begin{align}
Z_{\parallel,d} =\frac{1}{2\pi
c}&\left[\int_{S_B-S_A}(\nabla\varphi_{d,B})^2\,dS
-\int_{S_A}\nabla\varphi_{d,B}\cdot\nabla(\varphi_{d,A}-\varphi_{d,B})\,dS\right]\nonumber\\
=\frac{1}{2\pi c}&\left[\int_{C_B-C_A}\varphi_{d,B}{\bf
n}\cdot\nabla\varphi_{d,B}\,dl
-\int_{S_B-S_A}\varphi_{d,B}\nabla^2\varphi_{d,B}\,dS\right.\nonumber\\
&\left.-\int_{C_{A}}\varphi_{d,B}{\bf
n}\cdot\nabla(\varphi_{d,A}-\varphi_{d,B})\,dl
+\int_{S_{A}}\varphi_{d,B}\nabla^2(\varphi_{d,A}-\varphi_{d,B})\,dS
\right]\nonumber\\
&\!\!\!\!\!\!\!\!\!\!\!\!\!\!\!\!\!\!=-\frac{1}{2\pi
c}\int_{C_A}\varphi_{d,B}{\bf n}\cdot\nabla\varphi_{d,A}\,dl\ .
\end{align}
To go from the second to the third form of the impedance in this
equation, we use the fact that $\varphi_{d,B}=0$ on $C_B$, that
$\nabla^2\varphi_{d,B}=0$ in the region $S_B-S_A$, and that
$\nabla^2(\varphi_{d,A}-\varphi_{d,B})=0$ everywhere.

However, we can obtain formulas for the impedance of a step-out
transition that are even simpler. The dipole part of the impedance,
for example, can be written as
\begin{eqnarray}
Z_{\parallel,d} &=&\frac{1}{2\pi c}\left[\int_{C_B}\varphi_{d,B}{\bf
n}\cdot{\bf\nabla}\varphi_{d,B}\,dl
-\int_{S_{B}}\varphi_{d,B}\nabla^2\varphi_{d,B}\,dS\right.\nonumber\\
& &\quad\quad\left.-\int_{C_A}\varphi_{d,A}{\bf
n}\cdot{\bf\nabla}\varphi_{d,B}\,dl
+\int_{S_{A}}\varphi_{d,A}\nabla^2\varphi_{d,B}\,dS \right]\ .
\end{eqnarray}
The line integrals are zero again. Combining the remaining terms and
substituting for $\nabla^2\varphi$ from Eq.~(\ref{laplacemd_eq}), we
obtain
\begin{equation}
Z_{\perp,d}=\lim_{x,y\to 0}\frac{2}{\omega }\frac{\partial}{\partial
y}\left[\varphi_{d,B}(x,y)-\varphi_{d,A}(x,y)\right]\
.\label{Zd_out}
\end{equation}
Both dipole potential terms diverge in the limit $x,y\to0$; however,
the divergences are the same, and $Z_{\perp,d}$ is finite.

For the quad component of impedance of a step-out transition we
replace, in Eq.~(\ref{Rq_eq}), $S_{G}$ by $S_A$, and use Green's
first identity to obtain
\begin{equation}
Z_{\parallel,q} =\frac{1}{2\pi c}\left[
-\int_{S_{B}}\varphi_{q,B}\nabla^2\varphi_{m,B}\,dS +\int_{S_{A}}
\varphi_{m,A}\nabla^2\varphi_{q,B}\,dS\right]\ ,\label{Zqhelp_out}
\end{equation}
where the line integral terms are again zero. Substituting for the
Laplacians from Eqs.~(\ref{laplacemd_eq}), (\ref{laplaceq_eq}),
using the relation
$\lim_{x,y\to0}[\partial^2\varphi_{m,A}(x,y)/\partial
y^2-2\varphi_{q,A}(x,y)]=0$, and going to $Z_{\perp,q}$ we obtain
\begin{equation}
Z_{\perp,q} =\lim_{x,y\to
0}\frac{4}{\omega}\left[{\varphi_{q,B}(x,y)}-
{\varphi_{q,A}(x,y)}\right]\ .\label{Zq_out}
\end{equation}
Similarly, the transverse monopole impedance is
\begin{equation}
Z_{\perp,m}=\lim_{x,y\to
0}\frac{2}{\omega}\left[{\varphi_{d,B}(x,y)}-
{\varphi_{d,A}(x,y)}\right]\ ;\label{Zperpm_trans_eq}
\end{equation}
the longitudinal impedance is given by
\begin{equation}
Z_{\parallel,long} =\lim_{x,y\to
0}\frac{2}{c}\left[{\varphi_{m,B}(x,y)}-
{\varphi_{m,A}(x,y)}\right]\ .
\end{equation}
We see that, once the Green functions are known, obtaining the
impedances of step-out transitions (as well as long collimators) is
a relatively straightforward matter. Whereas an equation of the form
Eq.~(\ref{Z_trans_edif_eq}) can be interpreted to mean that the
longitudinal impedance in the optical regime is $2/c$ times the
static field energy of a charge distribution in region~$B$ minus
that of one in region~$A$, these simpler equations can be
interpreted to mean that the longitudinal impedance is $2/c$ times
the potential difference at the charge distribution in region~$B$
minus that at the charge in region~$A$.

As was the case with the irises/short collimators, for a transition
into a large beam pipe, one also doesn't need to know details of the
large beam pipe in order to calculate the leading order behavior of
a transverse impedance. In the equations for impedance one can take
the potentials in region~$B$ to be derived from the free-space Green
function, given in Eq.~(\ref{green_free_eq}).

\subsection*{T1. Flat Transition}

\begin{figure}[htbp]
\centering
\includegraphics*[height=50mm]{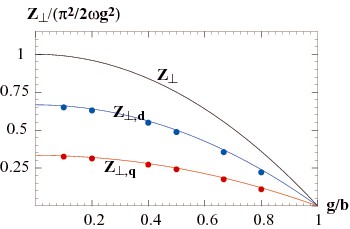}
\caption{For a step-out transition from a flat pipe of aperture $2g$
into a flat pipe of aperture $2b$, the transverse impedances
$Z_{\perp}$, $Z_{\perp,d}$, $Z_{\perp,q}$ as functions of
$\alpha=g/b$. Plotting symbols give ECHO numerical results for
comparison. }\label{flat_to_flat_trans_fi}
\end{figure}

In the case of a flat, symmetric, step-out transition going from
aperture $2g$ to $2b$, we substitute the potentials
Eqs.~(\ref{fi_flat_to_flat_eq}) into Eqs.~(\ref{Zd_out}),
(\ref{Zq_out}). We obtain
\begin{equation}
Z_\perp=\frac{\pi^2}{2\omega}\left(\frac{1}{g^2}-\frac{1}{b^2}\right)\
,
\end{equation}
with $Z_{\perp,q}=\frac{1}{2}Z_{\perp,d}$. We see thus that the
transverse impedance of a flat  step-out transition (or of a long,
flat collimator) is a factor $\pi^2/8$ times the transverse
impedance of a long, round collimator, if we take the half-heights
in the former case to be equal to the radii in the
latter~\cite{Palumbo}. In Fig.~\ref{flat_to_flat_trans_fi} we plot
the theoretical dependence and compare with ECHO numerical results
(the plotting symbols). We see that the agreement is very good.

If we perform the longitudinal impedance calculation for the flat
step-out transition we find that $Z_{\parallel,long}=4\ln(b/g)/c$,
which is the same as for the round case, if we take the
half-heights in the flat case to be equal to the radii in the
round one.

For a design orbit that is shifted by $\Delta y$ from the symmetry
axis, we find, using Eq.~(\ref{Zperpm_trans_eq}),
\begin{equation}
Z_{\perp,m}=\frac{\pi}{\omega}\left[\frac{1}{g}\tan\frac{\pi
\Delta y}{2g}-\frac{1}{b}\tan\frac{\pi \Delta y}{2b}\right]\ .
\end{equation}
Note that for a beam close to the wall at {\it e.g.} $y=g$,
$Z_{\perp,m}=2/[\omega(g-\Delta y)]$, which is twice as large is we
found for the close-to-the-wall impedance in the case of an iris
(see Eq.~(\ref{zm_iris_eq})).

\subsection*{T2. Rectangular Transition}

Consider a rectangular pipe of width $2w$ and height $2g$ that
transitions into a large beam pipe, and a design orbit that
follows the symmetry line of the rectangular pipe. The Green's
function for Poisson's equation in a rectangular pipe has been
obtained by Gluckstern, {\it et al}~\cite{Gluckstern}. Their
result is:\footnote{Note that there is a typo in their
Eq.~(5.11).}
\begin{align}
G(x,y,y_0)&=-4\sum_{n=1}^\infty\frac{e^{-\frac{n\pi w}{2g}}\cosh
\frac{n\pi x}{2g}}{n\cosh \frac{n\pi
w}{2g}}\sin\frac{n\pi}{2g}(y+g)\sin\frac{n\pi}{2g}(y_0+g)\nonumber\\
&+\ln\left\{\left[x^2+(y-y_0)^2\right] \frac{\sinh^2\frac{\pi
x}{4g}+\cos^2\frac{\pi}{4g}(y+y_0)}{\sinh^2\frac{\pi
x}{4g}+\sin^2\frac{\pi }{4g}(y-y_0)}\right\}\nonumber\\
&-\ln\left[{x^2+(y-y_0)^2}\right]\ .\label{rectangular_trans_eq}
\end{align}
The last term contains the singularity. The other terms are
everywhere finite, and the sum in the first term converges well.

\begin{figure}[htbp]
\centering
\includegraphics*[height=50mm]{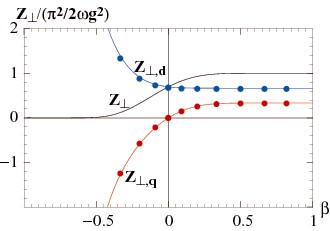}
\caption{For a rectangular step-out transition to a large beam pipe,
the leading order dependence of the transverse impedances
$Z_{\perp}$, $Z_{\perp,d}$, $Z_{\perp,q}$, normalized to
$\pi^2/(2\omega g^2)$, as functions of $\beta=(w-g)/(w+g)$. The
plotting symbols give ECHO numerical results for
comparison.}\label{rect_trans_fi}
\end{figure}

For our calculation we take region~$B$ to have the free space Green
function, {\it i.e.} the same as the last term in
Eq.~(\ref{rectangular_trans_eq}). Thus we have no singularities left
in our impedances, since they all involve the difference of
potentials in the two regions. Our final result is
\begin{eqnarray}
Z_{\perp,d}&=&\frac{\pi^2}{3\omega
g^2}\left[1+24\sum_{m=1}^\infty\frac{m}{1+e^{2\pi m\alpha}}\right]\
,\nonumber\\
 Z_{\perp,q}&=&\frac{\pi^2}{6\omega
g^2}\left[1-24\sum_{m=1}^\infty\frac{2m-1}{1+e^{\pi
(2m-1)\alpha}}\right]\ ,\label{Z_rect_trans_eq}
\end{eqnarray}
with $\alpha=w/g$. We perform the sums numerically. The results,
when normalized to $\pi^2/(2\omega g^2)$, are plotted as functions
of $\beta=(w-g)/(w+g)$ in Fig.~\ref{rect_trans_fi}
($Z_{\perp}=Z_{\perp,d}+Z_{\perp,q}$). Note that when $\beta= 1$ the
results agree with the leading order behavior for the flat
transition, given above. For $\beta=-1$ (a step-out transition from
an infinitely high vertical beam pipe)
$Z_{\perp,q}=-Z_{\perp,d}=-\pi^2/(6\omega w^2)$ and $Z_{\perp}=0$.
For the special case of a square beam pipe
$Z_{\perp}=Z_{\perp,d}=(0.697)\pi^2/(2\omega g^2)$, which is 86\% of
the result for a round pipe with radius $g$, $4/(\omega g^2)$; and
$Z_{\perp,q}=0$. The plotting symbols in the figure give ECHO
numerical results, and we see good agreement. Finally, note that the
horizontal impedance is obtained from Eqs.~(\ref{Z_rect_trans_eq})
by exchanging $w$ and $g$.

\subsection*{T3. Elliptical Transition}

Consider an elliptical pipe of horizontal axis $w$ and vertical axis
$g$ that transitions into a large beam pipe, and a design orbit that
follows the symmetry line of the elliptical pipe. The Green's
function for Poisson's equation in an elliptical pipe has been
obtained by Gluckstern, {\it et al}~\cite{Gluckstern}. For the case
$w\ge g$ it is given by \footnote{Note that there is a typo in their
version of this equation, their Eq. (4.36).}
\begin{align}
G(x,y,y_0)&=-4\sum_{n=1}^\infty\frac{e^{-nu_0}}{n}\left[\frac{{\rm
Re}T_n(\frac{x+iy}{d})\,{\rm Re}T_n(\frac{iy_0}{d})}{\cosh
nu_0}+\frac{{\rm Im}T_n(\frac{x+iy}{d})\,{\rm
Im}T_n(\frac{iy_0}{d})}{\sinh nu_0}\right]\nonumber\\
&-\ln\left[{x^2+(y-y_0)^2}\right]\ ,
\end{align}
with $T_n$ the Chebyshev polynomials of the first kind,
$d^2=w^2-g^2$, and $u_0={\rm arccoth}\, (w/g)$. [Note that $G$, as
given here, is not zero but constant on the elliptical boundary, a
fact, however, that does not affect our results.] The calculation
procedure is the same as for the rectangular step-out transition. In
the elliptical case we find that
\begin{align}
\hat \varphi_d(x,y)&=8\sum_{m=1}^\infty\frac{(-1)^m{\rm
Im}\left[T_{2m-1}(\frac{x+iy}{d})\right]}{e^{2(2m-1)u_0}-1}
,\nonumber\\
\hat \varphi_q(x,y)&=-4\sum_{m=1}^\infty\frac{(-1)^m{\rm
Re}\left[T_{2m}(\frac{x+iy}{d})\right]}{e^{4m u_0}+1} ,
\end{align}
with $\hat\varphi$ signifying the part of the potential that does
not contain the singularity. Note that the horizontal ($x$)
impedance of a step-out transition is equal to the (vertical)
impedance of the transition after it has been rotated by $90^\circ$.
Thus to obtain the impedances for an elliptical step-out transition
with $w<g$ we calculate the $x$ impedances for the rotated case,
following the analogous procedure to what we use for finding the $y$
impedances.

\begin{figure}[htbp]
\centering
\includegraphics*[height=50mm]{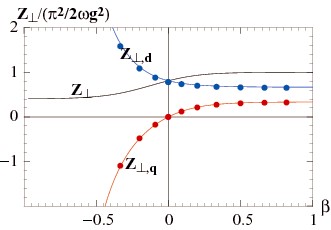}
\caption{For an elliptical step-out transition to a large beam pipe,
the leading order dependence of the transverse impedances
$Z_{\perp}$, $Z_{\perp,d}$, $Z_{\perp,q}$, normalized to
$\pi^2/(2\omega g^2)$, as functions of $\beta=(w-g)/(w+g)$. The
plotting symbols give ECHO numerical results for
comparison.}\label{ellipse_trans_fi}
\end{figure}

We finally obtain
\begin{align}
Z_{\perp,d}&=\frac{16}{\omega
g^2(\alpha^2-1)}\sum_{m=1}^\infty\frac{2m-1}{\left(\frac{\alpha+1}{\alpha-1}\right)^{2m-1}-1}\
,\nonumber\\
 Z_{\perp,q}&=\frac{32}{\omega
g^2(\alpha^2-1)}\sum_{m=1}^\infty\frac{m}{\left(\frac{\alpha+1}{\alpha-1}\right)^{2m}+1}\
,\label{ellipse_trans_eq}
\end{align}
with $\alpha=w/g$ (valid for all $\alpha$). We have simplified the
results by using the relation $e^{2u_0}=(\alpha+1)/|\alpha-1|$. The
sums are performed numerically. The results are plotted as functions
of $\beta=(w-g)/(w+g)$ in Fig.~\ref{ellipse_trans_fi}. The plotting
symbols in the figure give ECHO numerical results, and we see good
agreement. We see that we obtain the expected results for the round
case ($\beta=0$), $Z_{\perp,d}=4/(\omega g^2)$ and $Z_{\perp,q}=0$;
for the flat case ($\beta=1$),
$Z_{\perp,d}=2Z_{\perp,q}=\pi^2/(3\omega g^2)$. For the case of a
horizontally infinitesimally narrow elliptical pipe ($\beta=-1$) we
find that $Z_{\perp,d}\approx -Z_{\perp,q}=1/(\omega w^2)$, and that
the total impedance $Z_\perp=2/(\omega g^2)$. Note that in this
limit $Z_\perp$ has the same value as we obtained for any small
elliptical iris in a beam pipe (see Eq.~(\ref{ellipse_iris_eq})).
Details of how all three limits are arrived at can be found in the
Appendix.

\section{More Complicated Transitions}

\begin{figure}[htbp]
\centering
\includegraphics*[width=90mm]{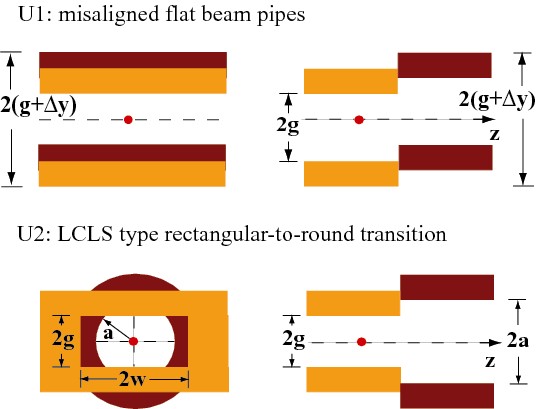}
\caption{Cross-section view (from upstream end; left figures) and
longitudinal view (right figures) of more complicated transitions:
U1--misaligned flat pipes, and U2--LCLS-type rectangular-to-round
transitions. Dimension labels are given; the design orbit location
is indicated by the red dot. Two colors are used as an aid in
visualization.}\label{complicated_geometries_fi}
\end{figure}

\begin{figure}[htbp]
\centering
\includegraphics*[height=35mm]{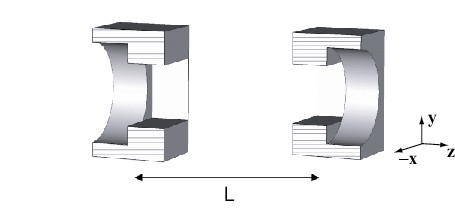}
\caption{A matching pair of LCLS rectangular-to-round transitions in
cut-away perspective view. The distance $L$ between transitions is
much larger than the catch-up distance for the nominal bunch
length.}\label{LCLS_3d_fi}
\end{figure}

In this section we give two examples that are neither an
iris/short collimator in a beam pipe, nor a step-in or step-out
transition. The examples are: (U1)--misaligned flat beam pipes,
and (U2)--LCLS-type rectangular-to-round transitions. See
Fig.~\ref{complicated_geometries_fi}. A cut-away perspective view
of a pair of the LCLS transitions is also given in
Fig.~\ref{LCLS_3d_fi}.

\subsection*{U1. Misaligned Flat Beam Pipes}

Consider first two flat beam pipes with thick walls and aperture
$2g$ that are perfectly aligned and joined at $z=0$. The design
orbit lies in the horizontal symmetry plane. Now imagine shifting
the $z>0$ pipe vertically by $\Delta y$ ($|\Delta y|<g$) and the
$z<0$ pipe by $-\Delta y$, and keeping the design orbit unchanged.
Note that the resulting transition no longer has a horizontal
symmetry plane.

Let us sketch out the calculation of the transverse impedance
$Z_{\perp,m}$ for this structure (for $\Delta y>0$). The
potentials for this problem are given by
\begin{align}
\varphi_{m,A}(x,y)&=G(x,y+\Delta y,\Delta y)\
,\quad\varphi_{d,A}(x,y)=\left.\frac{\partial G}{\partial
y_0}(x,y+\Delta y,y_0)\right|_{y_0=\Delta y}\ ,\nonumber\\
\varphi_{m,B}(x,y)&=G(x,y-\Delta y,-\Delta y)\
,\quad\varphi_{d,B}(x,y)=\left.\frac{\partial G}{\partial
y_0}(x,y-\Delta y,y_0)\right|_{y_0=-\Delta y}\ ,
\end{align}
with the flat pipe Green function $G(x,y,y_0)$ given by
Eq.~(\ref{green_parallel_eq}), but with $b$ replaced by $g$. Note
that, for this geometry, the aperture $S_G$ is the intersection of
$S_A$ and $S_B$.

\begin{figure}[htbp]
\centering
\includegraphics*[height=50mm]{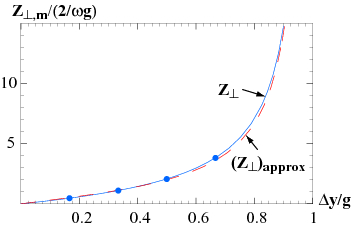}
\caption{For the misaligned, flat pipe of aperture $2g$, the
transverse impedance $Z_{\perp,m}$ as function of misalignment
parameter $\Delta y/g$. Plotting symbols give ECHO results. The
approximation, $(Z_{\perp,m})_{approx}$, is also shown
(dashes).}\label{misaligned_pipe_Ztrans_fi}
\end{figure}

Beginning with Eq.~(\ref{Rm_eq}), and using Green's identity, we
obtain
\begin{eqnarray}
Z_{\parallel,m}&=&\frac{1}{2\pi
c}\left[\int_{C_B}\varphi_{d,B}{\bf
n}\cdot\nabla\varphi_{m,B}\,dl-\int_{S_B}\varphi_{d,B}\nabla^2\varphi_{m,B}\,dS\right.\nonumber\\
&&- \left.\int_{C_G}\varphi_{d,B}{\bf
n}\cdot\nabla\varphi_{m,A}\,dl
+\int_{S_G}\varphi_{d,B}\nabla^2\varphi_{m,A} \,dS\right] \
.\label{misaligned_pipe_Zfirst_eq}
\end{eqnarray}
The first integral above is zero, because $\varphi_{d,B}$ is zero on
boundary $C_B$, and the second and fourth integrals cancel. We are
left with the third integral, which implies a transverse impedance
(see Eqs.~\ref{zperp_zparallel_eq}):
\begin{align}
Z_{\perp,m}&=-\frac{1}{2\pi \omega}\int_{C_G}\varphi_{d,B}{\bf
n}\cdot\nabla\varphi_{m,A}\,dl\nonumber\\
&=-\frac{1}{2\pi \omega}\int_0^\infty\varphi_{d,B}(x,g-\Delta
y)\frac{\partial\varphi_{m,A}}{\partial y}(x,g-\Delta y)\,dx\ .
\end{align}
Note that the contribution of the integral at $y=-g+\Delta y$ is
zero because this is a boundary of region $B$. We obtain, finally,
the analytical result (valid for either sign of $\Delta y$)
\begin{equation}
Z_{\perp,m}=\frac{1}{\omega g}\left[{\rm sgn}(\Delta
y)-\pi\left(1+\frac{|\Delta y|}{g}\right)\cot\frac{\pi\Delta
y}{g}+\pi\csc\frac{\pi\Delta y}{g}\right]\quad\quad[|\Delta y|<g]\
,\label{misaligned_pipe_Ztrans_eq}
\end{equation}
with ${\rm sgn}(x)$ meaning the sign of $x$. We note that
$Z_{\perp,m}$ is positive for positive $\Delta y$, and that it is
odd in $\Delta y$. Also, note that as $\Delta y\to g$,
$Z_{\perp,m}\to 3/[\omega(g-\Delta y)]$. The result of
Eq.~\ref{misaligned_pipe_Ztrans_eq} (for $\Delta y>0$) is plotted in
Fig.~\ref{misaligned_pipe_Ztrans_fi}. Plotting symbols give ECHO
results; we see excellent agreement with our results. The function,
\begin{equation}
(Z_{\perp})_{approx}=\frac{3}{\omega
g}\left(\frac{1}{1-\frac{\Delta y}{g}}-\frac{1}{1+\frac{\Delta
y}{g}}\right)\ ,
\end{equation}
gives a good approximation to the impedance (see the dashes in the
figure).

\begin{figure}[htbp]
\centering
\includegraphics*[height=50mm]{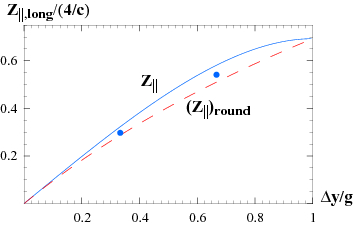}
\caption{For the misaligned, flat pipe of aperture $2g$, the
longitudinal impedance $Z_{\parallel,long}$ as function of
misalignment parameter $\Delta y/g$. Plotting symbols give ECHO
results. As a scale comparison the impedance of a round, step-out
transition, from radius $g$ to radius $g+\Delta y$,
$(Z_{\parallel})_{round}=4\ln(1+\Delta y/g)/c$, is also given
(dashes).}\label{misaligned_pipe_Zlong_fi}
\end{figure}

Obtaining the longitudinal impedance $Z_{\parallel,long}$ for the
misaligned pipe, one follows a similar procedure. In this case we
find that the solution is given by
\begin{align}
Z_{\parallel,long}&=-\frac{1}{\pi
c}\int_0^\infty\varphi_{m,B}(x,g-\Delta
y)\frac{\partial\varphi_{m,A}}{\partial
y}(x,g-\Delta y)\,dx\quad\quad\quad\quad[\Delta y>0]\nonumber\\
&=-\frac{2\cos\alpha}{\pi}\int_0^\infty\ln\left[\frac{\cosh
x-\sin\alpha}{\cosh x+\sin3\alpha}\right]\frac{dx}{\cosh
x-\sin\alpha}\ ,
\end{align}
with $\alpha=\frac{1}{2}\pi\Delta y/g$. We solve the integral
numerically. The result, for $\Delta y>0$, is plotted in
Fig.~\ref{misaligned_pipe_Zlong_fi}. Note that $Z_{\parallel,long}$
is even with respect to $\Delta y$. The plotting symbols give ECHO
results, and we see reasonably good agreement with our results. As a
scale comparison the impedance of a round, step-out transition, from
radius $g$ to radius $g+\Delta y$,
$(Z_{\parallel,long})_{round}=4\ln(1+\Delta y/g)/c$, is also given
in the plot (the dashes).

\subsection*{U2. LCLS Rectangular-to-Round Transition}

In the LCLS undulator region there are 33 pairs of
rectangular-to-round transitions. The rectangular aperture has
horizontal width $2w=10$~mm by vertical height $2g=5$~mm; the
round aperture has radius $a=4$~mm. The axes of the two pipes are
aligned. The transitions are abrupt. The bunch length in this
region is 20~$\mu$m (rms). Thus our optical regime formulas are
applicable. Note that these transitions are neither step-in nor
step-out transitions. In the LCLS undulator region the
longitudinal impedance is the more important one, and this is the
one we calculate here. A numerical ECHO calculation has found that
the longitudinal impedance of a pair of transitions (one
rectangular-to-round and one round-to-rectangular transition)
equals $1.21/c$~\cite{BaneZ06}.

Beginning with Eq.~(\ref{Rlong_eq}), and using Green's first
identity, we obtain
\begin{equation}
Z_{\parallel,long}=-\frac{1}{2\pi c}\int_{C_G}\varphi_{m,B}{\bf
n}\cdot\nabla\varphi_{m,A}\,d\ell\ .
\end{equation}
Here $S_G$ is the intersection of $S_A$ and $S_B$. The integration
path $C_G$ follows the rectangular boundary at the top and bottom,
and the circular boundary on the right and left (see
Fig.~\ref{complicated_geometries_fi}). In the rectangular-to-round
case [or we can say the rectangular-to-circular (rtc) case] with
impedance $Z_{\parallel,rtc}$ region~$A$ is the rectangular pipe,
region~$B$ is the circular pipe. In the circular-to-rectangular
(ctr) case with impedance $Z_{\parallel,ctr}$ the regions are
reversed.

The circular monopole potential is given by
$\varphi_{m,c}=-\ln[(x^2+y^2)/a^2]$; and the rectangular monopole
potential, by $\varphi_{m,r}=G(x,y,0)$, with $G$ given by
Eq.~(\ref{rectangular_trans_eq}). Then we have as
rectangular-to-circular impedance
\begin{equation}
Z_{\parallel,rtc}=-\frac{2}{\pi
c}\int_{0}^{\sqrt{a^2-g^2}}\!\!\!\!\!\!\!\!\varphi_{m,c}(x,g)\frac{\partial\varphi_{m,r}}{\partial
y}(x,g)\,dx\ .
\end{equation}
The contribution from the circular part of the boundary is zero,
since $\varphi_{m,c}$ is zero on this boundary. In the
circular-to-rectangular case
\begin{equation}
Z_{\parallel,ctr}=-\frac{2a}{\pi
c}\int_{0}^{\arctan(g/\sqrt{a^2-g^2})}\!\!\!\!\!\!\!\!\!\!\varphi_{m,r}(a\cos\theta,a\sin\theta)\frac{\partial\varphi_{m,c}}{\partial
r}(a\cos\theta,a\sin\theta)\,d\theta\ ,
\end{equation}
where $\partial\varphi_{m,c}/\partial r(x,y)=-2/\sqrt{x^2+y^2}$. The
contribution from the rectangular part of the boundary is zero,
since $\varphi_{m,r}$ is zero on this boundary. These integrals can
easily be solved numerically, and the sums coming from
$\varphi_{m,r}$ converge well.

\begin{figure}[htbp]
\centering
\includegraphics*[height=50mm]{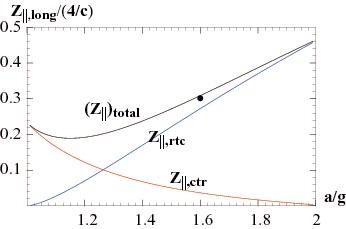}
\caption{Longitudinal impedance for transitions of the LCLS
rectangular-to-circular type, giving $Z_{\parallel,rtc}$,
$Z_{\parallel,ctr}$, and their sum $(Z_{\parallel})_{total}$ as
functions of circular radius $a$. The rectangle width $2w=4g$. The
ECHO result for $(Z_{\parallel})_{total}$, from Ref.~\cite{BaneZ06},
is given by the black dot.} \label{lcls_fi}
\end{figure}

To do a small parameter study, let us keep the shape of the
rectangular pipe fixed, with $w=2g$, and let us vary $a$ from $g$ to
$w$. In Fig.~\ref{lcls_fi} we plot the results, giving the impedance
of a rectangular-to-circular transition, $Z_{\parallel,rtc}$, of a
circular-to-rectangular transition, $Z_{\parallel,ctr}$, and the sum
of one of each type, $(Z_{\parallel})_{total}$, as functions of
$a/g$. We see that each of the single transition curves goes to zero
exactly where that transition becomes a step-in transition. For the
actual design of the LCLS transition ($a/g=1.6$), the $rtc$
transition has 7.5 times the impedance of the $ctr$ transition. In
Fig.~\ref{lcls_fi} the black dot gives $(Z_{\parallel})_{total}$ as
obtained by ECHO, and we see good agreement: our calculation gives
$1.24/c$ and the ECHO result is $1.21/c$.

\section{Conclusions}

We have used a method, that we derived in a companion
report~\cite{Stupakov07}, to find impedances in the optical regime,
and applied it to various 3D beam pipe transitions that one
encounters in vacuum chambers of accelerators. The method is
applicable to high frequencies and transitions that are short
compared to the catch-up distance. Our examples are of four types:
an iris/short collimator in a beam pipe, a step-in transition, a
step-out transition, and more complicated transitions. (Note that a
long collimator with ends that are short transitions has an
impedance that is the sum of the impedances of a step-in and a
step-out transition, and is thus also included.) Most of our results
are analytical, with a few given in terms of a simple one
dimensional integral. We believe that all of our results are new. We
have also compared (most of) our results with numerical simulations
with the computer program ECHO, a finite-difference program that
solves Maxwell's equations of an ultra-relativistic bunch within
metallic boundaries of 3D geometry, and the agreement is excellent.
Note that our method is a much simpler way of obtaining impedances
than the simulations.

We have focused on transverse impedances. For bi-symmetric
(horizontal and vertical mirror symmetric) examples, a bunch moving
at a small offset from the symmetry axis will excite a transverse
impedance composed of both a dipole and a quadrupole component. For
such problems we give both components. The
iris/short-collimator-in-a-beam-pipe examples we solve include
irises with small aperture that are (I1)~flat, (I2)~rectangular,
(I3)~elliptical; also included is (I4)~a flat iris (not necessarily
small) in a flat beam pipe. An interesting result is that the
vertical impedance of an elliptical iris is independent of the
horizontal axis of the ellipse.

For a step-in transition of any shape we find that all impedances
(transverse and longitudinal) are zero. For a step-out transition
(which in the optical regime has the same impedance as a long
collimator) we give the solution for (T1)~a flat step-out transition
to a flat beam pipe, (T2)~a rectangular step-out transition, and
(T3)~an elliptical step-out transition. We find, for example, that
the transverse impedance of a flat, long collimator in a flat beam
pipe is $\pi^2/8$ times the impedance of the inscribed round, long
collimator in a round beam pipe. The more complicated transitions
examples we solve are (U1)~misaligned flat pipes and (U2)~the LCLS
rectangular-to-round transitions.

The method of Ref.~\cite{Stupakov07} is powerful; it allows one to
calculate the impedance in the optical regime of a truly large class
of transitions. We have demonstrated this with a small number of
relatively simple examples, compared to what is possible.

\section*{Acknowledgements}
We thank  CST GmbH for letting us use CST MICROWAVE STUDIO for the
meshing for the ECHO simulations. We also thank B.~Podobedov for
running a GdfidL test simulation for us.

\section*{Appendix: Limiting Values of the Impedance of Elliptical Transitions}

The solution for the elliptical step-out transition with axes $w$ by
$g$ (horizontal by vertical), Eqs.~\ref{ellipse_trans_eq}, can be
written as
\begin{align}
Z_{\perp,d}&=\frac{4(1-\beta)^2}{\omega
g^2\beta}\sum_{m=1}^\infty\frac{2m-1}{\beta^{-(2m-1)}-1}\
,\nonumber\\
 Z_{\perp,q}&=\frac{8(1-\beta)^2}{\omega
g^2\beta}\sum_{m=1}^\infty\frac{m}{\beta^{-2m}+1}\ ,
\end{align}
where $\beta=(w-g)/(w+g)$. We derive here the limits for $\beta=0$
(a step-out transition from a round beam pipe), $\beta=1$ (from an
elliptical pipe with infinitesimal height), and $\beta=-1$ (from an
elliptical pipe with infinitesimal width).

\subsubsection*{The Limit for $\beta=0$}
In this limit, only the $m=1$ term contributes to $Z_{\perp,d}$ and
no term contributes to $Z_{\perp,q}$. We obtain the round step-out
transition results
\begin{equation}
Z_{\perp,d}=\frac{4}{\omega g^2}\ ,\quad\quad Z_{\perp,q}=0\ .
\end{equation}

\subsubsection*{The Limit for $\beta=1$}
Let us consider the dipole impedance first. For $\beta=1-\epsilon$,
with $\epsilon$ a small positive number, the sum peaks when
$2m\epsilon\approx\ln m$, {\it i.e.} for a large value of $m$. Thus
the sums can be replaced by integrals, $\sum_m\to\int dm$. Changing
variables to $x=\beta^{-(2m-1)}$, $dx=-2(\ln\beta) x\,dm$, we obtain
\begin{equation}
Z_{\perp,d}=\lim_{\beta\to1}\frac{2}{\omega
g^2}\frac{(1-\beta)^2}{\beta(\ln\beta)^2}\int_1^\infty\frac{\ln
x}{(x-1)}\frac{dx}{x}\ .
\end{equation}
The integral equals $\pi^2/6$, and
\begin{equation}
\lim_{\beta\to1}\frac{(1-\beta)^2}{\beta(\ln\beta)^2}=1\ .
\end{equation}
The calculation for $Z_{\perp,q}$ is similar. Our final result is
the same as for the flat pipe step-out transition
\begin{equation}
Z_{\perp,d}=\frac{\pi^2}{3\omega g^2}\ ,\quad\quad
Z_{\perp,q}=\frac{\pi^2}{6\omega g^2}\ ,\quad\quad
Z_{\perp}=\frac{\pi^2}{2\omega g^2}\ .
\end{equation}

\subsubsection*{The Limit for $\beta=-1$}

Let $\gamma=-\beta$. Then our equations become
\begin{align}
Z_{\perp,d}&=\frac{4(1+\gamma)^2}{\omega
g^2\gamma}\sum_{m=1}^\infty\frac{2m-1}{\gamma^{-(2m-1)}+1}\
,\nonumber\\
 Z_{\perp,q}&=-\frac{8(1+\gamma)^2}{\omega
g^2\gamma}\sum_{m=1}^\infty\frac{m}{\gamma^{-2m}+1}\ .
\end{align}
To find the leading order behavior of $Z_\perp$ we need to go to
second order in the calculation. To do this we will use the
Euler-Maclaurin formula relating sums to integrals~\cite{wiki}
\begin{equation}
\sum_{m=1}^{\infty}f(m)\approx \frac{f(1)}{2}+\int_1^\infty
f(x)\,dx-\frac{1}{12}f'(1)\ ,\label{Euler_eq}
\end{equation}
where $'$ denotes taking the derivative of a function. The formula
is valid if the sum converges. The approximation is good if
$\int_1^\infty |f{'''}(x)|\,dx$ is small.

Let us consider first the dipole part. We want the solution for
$\gamma=1-\epsilon$, with $\epsilon$ a small, positive parameter
that we, in the end, let go to zero. For the dipole part
\begin{equation}
f(m)=\frac{16}{\omega g^2}\frac{2m-1}{\gamma^{-(2m-1)}+1}\ .
\end{equation}
There are three terms on the right side of Eq.~\ref{Euler_eq} that
we need to calculate. The first term $f(1)/2=4/(\omega g^2)$. The
integral term, done like before, is
\begin{equation}
\int_{1}^\infty f(x)\,dx=\lim_{\gamma\to1}\frac{8}{\omega
g^2}\frac{1}{(\ln\gamma)^2}\int_{1/\gamma}^\infty\frac{\ln
x}{(x+1)}\frac{dx}{x}\ .\label{help_appendix_eq}
\end{equation}
Note that for this order of calculation, in the integral on the
right, the lower limit of integration is $1/\gamma$ instead of 1.
This integral
\begin{equation}
\int_{1/\gamma}^\infty\frac{\ln
x}{(x+1)}\frac{dx}{x}=\frac{\pi^2}{6}-\ln\left(\frac{1+\gamma}{\gamma}\right)
\ln{\gamma}-\frac{1}{2}(\ln{\gamma})^2+{\rm Li}_2(-1/\gamma)\ ,
\end{equation}
with ${\rm Li}_2(x)$ the polylogarithmic function of order 2. We
combine this result with the terms in front of the integral in
Eq.~\ref{help_appendix_eq}, expand around $\gamma=1$, and substitute
$\gamma=(g+w)/(g-w)$; we find that the second, integral contribution
to Eq.~\ref{Euler_eq} equals
\begin{displaymath}
\frac{1}{\omega}\left[\frac{\pi^2}{6w^2}+\left(\frac{\pi^2}{18}-2\right)\frac{1}{g^2}\right]\
.
\end{displaymath}

The third term in Eq.~\ref{Euler_eq}, $-f'(1)/12=-4/(3g^2)$. We sum
all three contributions to obtain $Z_{\perp,d}$. Exactly the same
technique is used for $Z_{\perp,q}$ (note, however, that the
equation corresponding to Eq.~\ref{help_appendix_eq} will have an
integral with lower limit $1/\gamma^2$, not $1/\gamma$). We finally
obtain
\begin{align}
Z_{\perp,d}&=\frac{1}{\omega}\left[\frac{\pi^2}{6w^2}+\left(\frac{2}{3}
+\frac{\pi^2}{18}\right)\frac{1}{g^2}\right]\nonumber\\
Z_{\perp,q}&=\frac{1}{\omega}\left[-\frac{\pi^2}{6w^2}+\left(\frac{4}{3}
-\frac{\pi^2}{18}\right)\frac{1}{g^2}\right]\ .
\end{align}
The leading order behavior of these impedances, in this limit, is
$Z_{\perp,d}=-Z_{\perp,q}=\pi^2/(6\omega w^2)$. Summing the two
impedances together, we find that the total impedance equals
$Z_\perp=2/(\omega g^2)$. This value of total impedance is the same
as we found for {\it any} small elliptical iris in a beam pipe. This
result appears to agree with the numerical calculation of the
original sums, which is plotted in Fig.~\ref{ellipse_trans_fi}.

\end{document}